\documentclass[12pt]{article}
\usepackage{osajnl2}
\usepackage[scriptsize]{caption}
\usepackage{amssymb}
\usepackage{subfigure}
\usepackage{braket}
\usepackage{amsmath}

\newcommand{\da}[1]{#1^{\dagger}}
\usepackage{color}

\begin{document}
\title{Quantum random walks with multiphoton interference and high-order correlation functions}

\author{Bryan T Gard$^{1,*}$, Robert M Cross$^1$, Petr M Anisimov$^1$, Hwang Lee$^1$ and Jonathan P Dowling$^{1,2}$}
\date{\today}

\address{\small{$^1$ Hearne Institute for Theoretical Physics, Department of Physics and Astronomy, Louisiana State University, 202 Nicholson Hall,  Baton Rouge, LA 70803, USA \\
$^2$ Computational Science Research Center, No.3 HeQing Road, Haidian District, Beijing, 100084, China} }

\email{Corresponding author$^*$: bgard1@lsu.edu}

\begin{abstract}
We show a simulation of quantum random walks with multiple photons using a staggered array of 50/50 beam splitters with a bank of detectors at any desired level.  We discuss the multiphoton interference effects that are inherent to this setup, and introduce one, two, and threefold coincidence detection schemes. The use of Feynman diagrams are used to intuitively explain the unique multiphoton interference effects of these quantum random walks. \\
\end{abstract}

\maketitle
\section{Introduction}
Random walks are ubiquitous in fields like physics, chemistry, biology, and economics.  In a generic version of a random walk with discrete steps, every time a walker arrives at a crossroad, he has to choose the route to take. After several crossings and choices, made for example by flipping a coin [with heads (tails) leading to a step to the left (right)], he will have followed one out of many possible paths. For a quantum walker, in contrast, the result of each coin toss is a superposition of heads and tails. That is, the quantum coin takes both states simultaneously and therefore the walker follows all the possible paths \cite{aharonov, kempe}.

The results of a classical walk are intuitive and relatively easy to comprehend. Quantum random walks, however, involve superpositions and therefore show strange unintuitive results. This effect causes the possibility for the walker to cause interference with himself, and either increase his probability to be at this crossing or decrease his probability by destructively interfering with himself. There is an increasing interest to understand quantum counterparts of classical random walks and to find out whether quantum effects are already exploited by nature. It has been suggested that single photon random walks have applications for quantum algorithms\cite{shenvi,lovett,farhi,perets} and for quantum computation \cite{kendon,trav,du}. Additionally, continuous time quantum random walks with two identical particles have been shown to solve the graph isomorphism problem \cite{gamble}. However,  most previous experimental realizations have been limited to single-particle quantum walks \cite{sanders,dur}, which 
have an exact mapping to classical wave phenomena \cite{knight}, and therefore cannot provide any advantage (computational or otherwise) as a result of uniquely quantum mechanical behavior.  Recently however, some experiments have used schemes similar to ours, with multiple photons\cite{crespi,crespi2,spring,tillmann}. Other recent experiments have investigated the properties of ``four-sided coin" random walks using two trapped ions \cite{zahringer}. It is important to note that in contrast to some of these references, which describe the random walk using a position and coin degree of freedom, we choose to adopt a description more closely related to experiment implementations (``coinless") and describe our walk only using paths and beam splitters.

Discrete quantum random walks with photons can be achieved by using a staggered array of 50/50 beam splitters each representing a step left or right. Now two walkers realized as photons were observed to pass through the same optical path network. For the first time, the two indistinguishable walkers were shown to interfere with each other \cite{peruzzo}. In contrast, for quantum walks of more than one indistinguishable particle, classical theory no longer provides a sufficient description. Quantum theory predicts that probability amplitudes interfere, leading to distinctly non-classical correlations \cite{omar}.

In this paper, we show that there are interesting aspects to these discrete, higher level, multi-photon random walks. Dirac's dictum, that every photon only interferes with itself is wrong. Hong, Ou, and Mandel showed that two photons incident on a beam splitter tend to interfere and both go one way and both go the other way\cite{hom}. However, the outcomes of these higher level multi-photon random walks are harder to visualize. We investigate onefold, twofold, and threefold probability coincidence detections in order to extract quantum information about the walks with multiple walkers from the output of these random walks. 
Based on ``Feyman diagrams'', we show that although quantum random walks run in parallel, coincidence detection at a particular site responds to specific paths only.

In future research, we plan to explore higher order correlations, use more generalized number resolving detection schemes, and investigate genetic algorithms in order to steer photons using phase shifters at key points within the array to create desired outputs.
\section{Quantum random walks}
\subsection{Model}
\label{sec:model}
\begin{figure}[htbp]
  \centering
  \subfigure[] {\includegraphics[height=4cm]{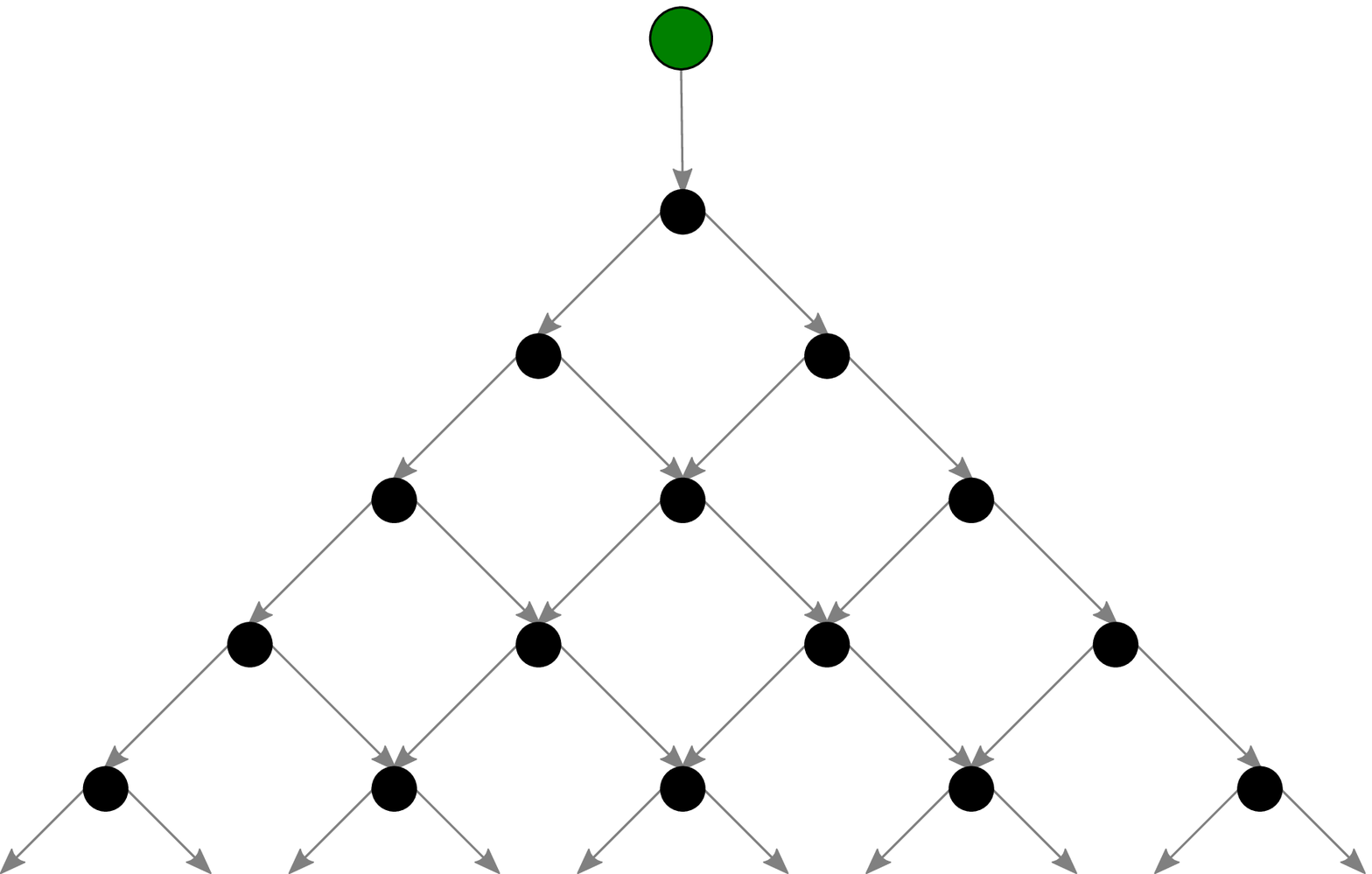}
  \label{fig:classical}
  }\hspace{5pt}
  \subfigure[]{\includegraphics[height=4cm]{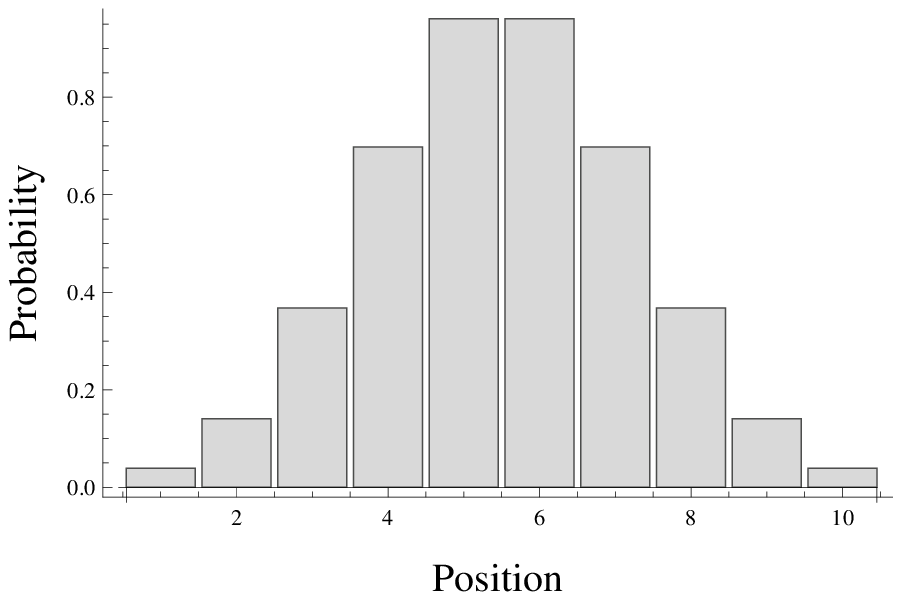}
  \label{fig:Gaussian}
  }
  \caption{(a) Classical random walk using a peg board and ping pong balls. (b) Result of a classical random walk is a binomial distribution. The maximum of the binomial distribution corresponds to the location directly below the starting position.}
\end{figure}

A peg board and a ping-pong ball, as shown in  Figure \ref{fig:classical}, is an example of a classical random walk, where the ping-pong ball is the walker that bounces left or right off the pegs on the board. Repeated walks through this setup result in outcomes that are distributed according to the binomial probability distribution with maximum probability right beneath the input peg as shown in Figure \ref{fig:Gaussian}.

\begin{figure}[bt]
  \centering
  \subfigure[] {\includegraphics[height=3cm]{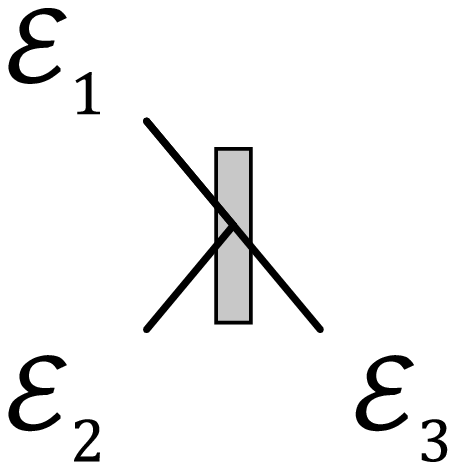}
  \label{fig:classicalsplitter}
  }
\hspace{1 cm}
  \subfigure[] {\includegraphics[height=3cm]{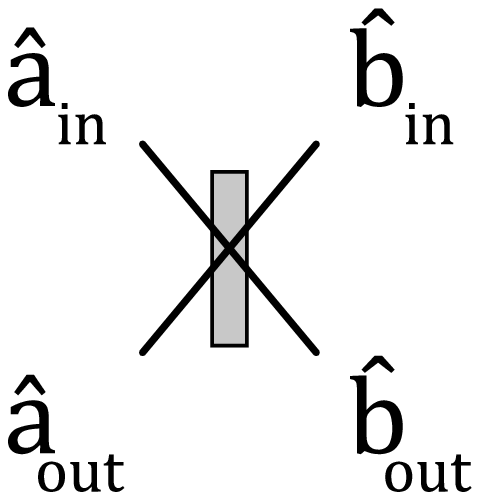}
  \label{fig:beamsplitter}
  }
  \caption{(a) Classical description of a beam splitter;
    (b) Quantum mechanical description of a beam splitter with the input and output field operators shown.}
\end{figure}

In the case of quantum random walks \cite{aharonov,kempe,knight,peruzzo,omar,mack,jeong}, we replace the pegs with 50/50 beam splitters and the ping-pong balls with photons. From a classical stand point they operate by splitting incident light so that half of the incident energy is reflected with an additional $\pi/2$ phase shift acquired by the field from the beam splitter, and half is transmitted through the beam splitter (see Figure \ref{fig:classicalsplitter}). 

There are many ways to model a beam splitter quantum mechanically. Yet, we have chosen to use the symmetrical model, which follows the physical properties of laboratory, optical, beam splitters, instead of an asymmetric, non-physical Hadamard model. This way, our results can be easily reproduced in an all-optical experiment. A quantum-mechanical description, adopted here, of such an experimental beam splitter is given by the transformation for the field operators illustrated in Figure \ref{fig:beamsplitter}, \cite{gerry},

\begin{equation}
  \left.\begin{aligned}
    {\hat{a}}^{\dag}_{in} &= \frac{1}{\sqrt2}(\textrm{i} {\hat{a}}^{\dag}_{out} +   {\hat{b}}^{\dag}_{out}), \\
    {\hat{b}}^{\dag}_{in} &=  \frac{1}{\sqrt2}({\hat{a}}^{\dag}_{out} + \textrm{i} {\hat{b}}^{\dag}_{out})    .
  \end{aligned} \right.
  \label{eq:creationtransformations}
\end{equation}

The terms that contain the factor of $i$ come from the fact that the incident photon is reflected off the beam-splitter instead of being transmitted through it, which gives the photon an additional $\pi/2$ relative phase \cite{gerry}. Here individual photons  after interactions with a beam-splitter end up in a superposition of being simultaneously transmitted and reflected. Consequently, the probability distributions for the outcomes are no longer binomial distributions, but multiphotons demonstrate quantum interference effects instead.

 We start with a two-mode quantum state of light with $N$ photons at the left port of a beam-splitter and $M$ at the right, written, $\ket{N,M} = \frac{1}{\sqrt{N!M!}} {\hat{a_0}}^{\dagger N} {\hat{b_0}}^{\dagger M} \ket{0_a,0_b}$. Here the input photons correspond to the creation operators, which after the first beam-splitter become

\begin{figure}[bt]
  \centering
  \includegraphics[height=4cm]{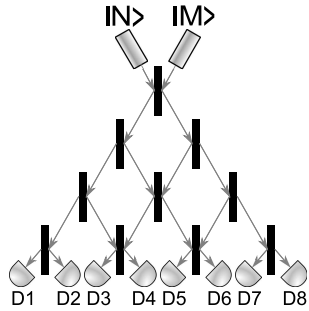}
  \caption{Pyramid structure of quantum random walk model using beam-splitters.}
  \label{fig:model}
\end{figure}

\begin{equation}
 \ket{N,M} = \frac{{\hat{a_0}}^{\dagger N} {\hat{b_0}}^{\dagger M}}{\sqrt{N!M!}}  \ket{0,0}
 \overset{\text{Eq. \ref{eq:creationtransformations}}}{\Longrightarrow}
 \ket{N,M}=\frac{1}{\sqrt{N!M!}} \left (\frac{\textrm i \da {\hat{a_1}}  +  \da {\hat{b_1}}}{\sqrt 2}\right)^N   \left (\frac{ \da {\hat{a_1}} + \textrm i \da {\hat{b_1}}}{\sqrt 2} \right)^M \ket{0_a,0_b},
\label{eq:input}
\end{equation}
where creation operators $\da {\hat{a_1}}$ and $\da {\hat{b_1}}$ corresponds to the photon exiting the left and right output port, respectively, of the beam-splitter \cite{gerry}.
This process is cascaded down through additional beam-splitters into a pyramid shape as shown in Figure \ref{fig:model}.
Hence, in order to propagate photons through the system, we apply the beam-splitter transformation, equation \ref{eq:creationtransformations} recursively onto  $\da {\hat{a}}$ and $\da {\hat{b}}$, while adding new creation operators that correspond to photons exiting from output ports of these beam-splitters, until we have reached the desired level. At this point, we turn the result from operator into the state notation by acting with the creation operator on the vacuum state, $ {\hat{a}}^{\dag n}\ket 0 = \sqrt{n!} \ket{n}$. As we propagate the state to the next level, we simply apply another beam-splitter transformation on each creation operator in Eq. \ref{eq:input}. Since we can apply the beam-splitter transformation freely inside of the parentheses of Eq. \ref{eq:input} and apply the powers $N$ and $M$ after we have fully propagated to a desired level, it is helpful to note that we need only fully propagate the states $\ket{1,0}$ and $\ket{0,1}$ in our system. Then the general case of an $\ket{N,M}$ 
state is just an expansion of order $N,M$ of the previously calculated operators from the $\ket{1,0}$ and $\ket{0,1}$ states. This simplification saves us a considerable amount of calculation time but while the case of $\ket{1,0}$ at level $L = 16$ can be calculated relatively easily, expanding the resulting operators for $N=M=9$ (that is $\ket{9,9}$) still presents computational issues.

\subsection{Detection of Outcomes: Correlation Functions}
\label{sec:Correlation}
The outcome of classical random walk is a walker exiting at a certain port. In a quantum random walk, the outcome is a state that describes the statistics of possible outcomes.
In order to recover this state, one can use conventional photodiodes. However, they do not distinguish the number of photons that actually hit them. In particular, the signal from a detector at position $m$ is proportional to the mean photon number $\bra{\psi}\da{\hat{a}_m}  {\hat{a}_m} \ket{\psi}$.
In single-photon experiments, the mean photon number is equal to the onefold probability that detector $m$ receives a photon. The probability distribution over all detectors provides all the information. It is important to note that single-photon and multi-photon walks are described by identical probability distributions. In experiments with multiple quantum walkers however, the onefold detection scheme throws away information, as it does not measure the exact number of photons that impact the detector and ignores what happens to other detectors.

Quantum random walks with multiple photons have unambiguous differences in the characteristics of the outcomes due to the interference between the two or more indistinguishable walkers. This new property of two, indistinguishable, interfering photons cannot be revealed with onefold probability.
The twofold probability expands upon the onefold probability, where the twofold probability at detectors $m$ and $n$ is equal to $\bra\psi \da {\hat{a}_m} \da {\hat{a}_n} {\hat{a}_n} {\hat{a}_m} \ket \psi$ \cite{rohde,bromberg,peruzzo}. The twofold probability reveals the correlations between simultaneous events at detectors $m$ and $n$ or acknowledges that two photons were received by the same detector. The twofold probability detection scheme is described by the joint probability that two detectors fire at the same time. This means that the twofold joint probability at detectors $m$ and $n$ represents the probability that detectors $m$ and $n$ each measure at least one photon simultaneously. Due to symmetry, the twofold probability at detectors $m$ and $n$ is equal to the twofold probability at detectors $n$ and $m$.

In the case where we have detector $m=n$, the twofold probability at detectors $m$ and $m$ represents the joint probability that detector $m$ fires and detector $m$ fires. We interpret this as detector $m$ receiving exactly two photons. After dividing by two due to the bosonic nature of photons, this interpretation holds well for two walkers and the twofold probability can be measured with conventional photo-diodes by repeating the same walk multiple numbers of times and counting the frequencies of $n,m$ events. Similar to the onefold probability, twofold probability throws away any information that involves three or more detectors firing simultaneously and it also can not distinguish between the cases where two photons impact a detector, or the cases where greater than two photons impact a detector.

Onefold and twofold probabilities have been thoroughly studied previously, \cite{peruzzo,rohde,bromberg,karski} and are mentioned in this paper for completeness. It is only logical for multi-walker quantum random walks to consider higher-order correlations. We present a case of threefold probability, which has not been previously studied, and discuss the visualization of the three dimensional results. It is defined similarly to lower-order probabilities; threefold probability at detectors $m$, $n$, and $l$ is equal to $\bra\psi \da {\hat{a}_m} \da {\hat{a}_n} \da {\hat{a}_l} {\hat{a}_l} {\hat{a}_n}  {\hat{a}_m} \ket \psi$. The threefold probability at detectors $m$, $n$, and $l$ represents the probability of measuring a photon at detectors $m$, $n$, and $l$ simultaneously. The threefold probability also has symmetry similar to the twofold probability, as the threefold probability at detectors $m$, $n$, and $l$ is equal to the threefold probability at detectors $n$, $l$, and $m$, and these are equal to any 
other permutations of $m$, $n$, and $l$ as well.

\section{Results}
\subsection{Onefold probability}

\begin{figure}[htbp]
  \centering
    \subfigure[] {\includegraphics[height=5.25cm]{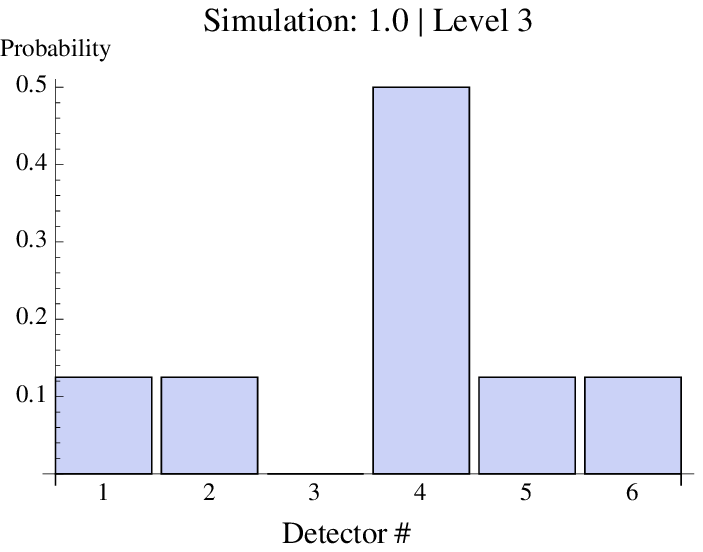}
    \label{fig:1foldplot}
    }
    \subfigure[] {\includegraphics[height=5.25cm]{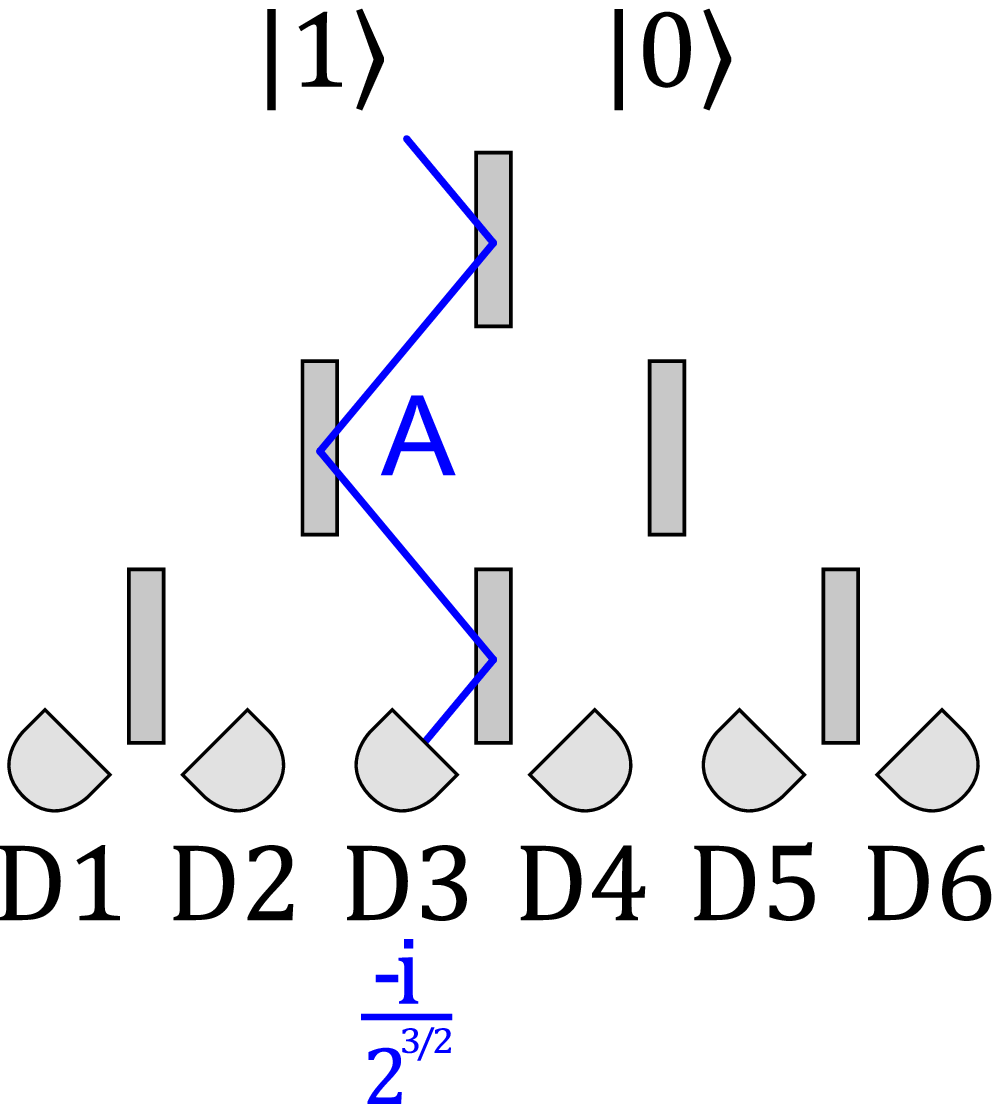}
    \label{fig:1foldfeynman1}
    }
    \subfigure[] {\includegraphics[height=5.25cm]{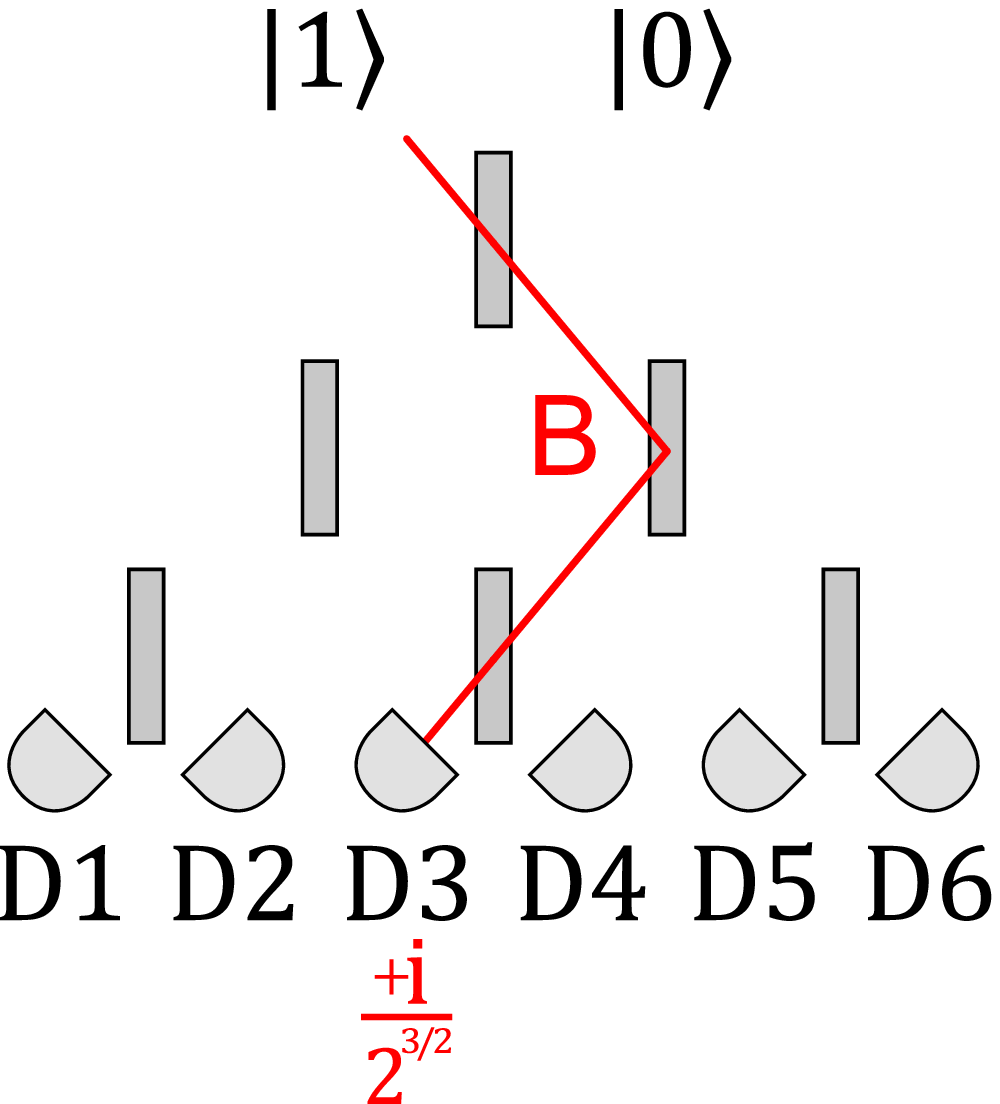}
    \label{fig:1foldfeynman2}
    }
    \subfigure[] {\includegraphics[height=5.25cm]{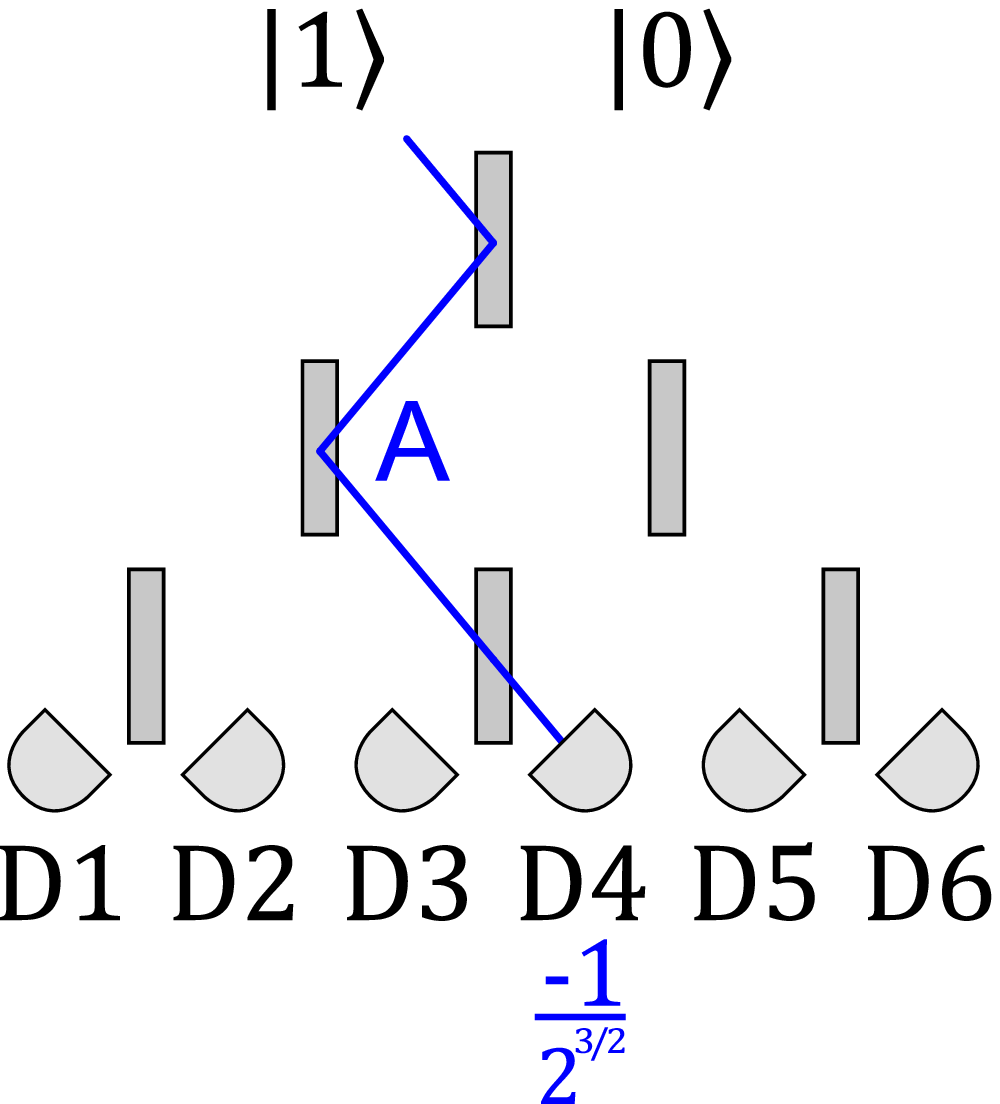}
    \label{fig:1foldfeynman3}
    }
    \subfigure[] {\includegraphics[height=5.25cm]{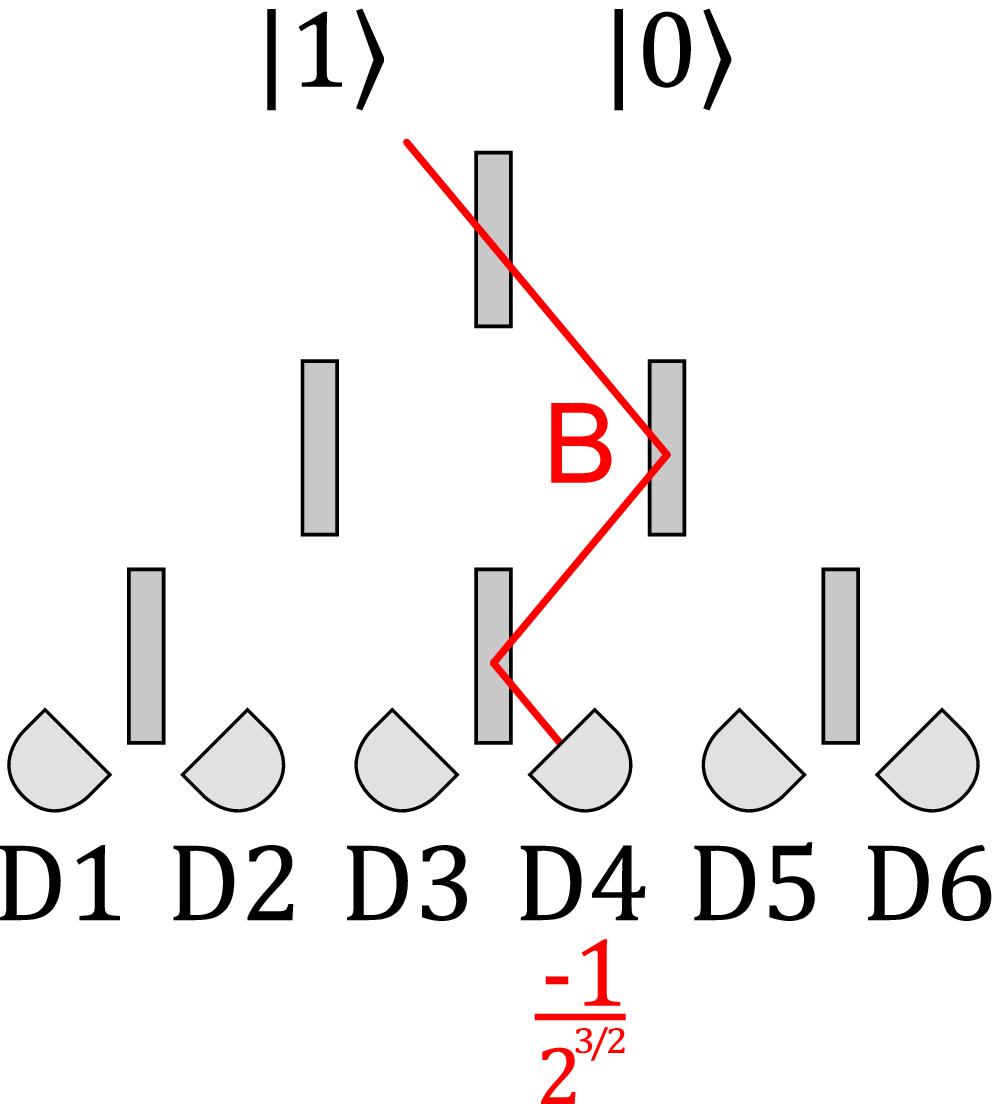}
    \label{fig:1foldfeynman4}
    }
\caption{(a) Plot of onefold probabilities showing single photon interference of $\ket{1,0}$ input state propagated through three levels of beam-splitters. Complete destructive interference is shown at detector D3 while constructive interference occurs at detector D4. Two Feynman diagrams required to explain the complete destructive interference at detector D3 (b,c). Feynman diagrams explaining the constructive interference at detector D4 (d,e). Only one path is shown per diagram since there is only one photon in the system.}
\label{plots}
\end{figure}

As it was pointed out in Section \ref{sec:Correlation}, onefold probabilities with more than one photon in the system produce similar results without providing new information. Because of this, we only look at input states with one photon, $\ket{1,0}$. A flipped input state, $\ket{0,1}$, results in a flipped output state, due to our choice for symmetrical beam-splitters. Figure \ref{fig:1foldplot} shows the onefold probabilities of a $\ket{1,0}$ input state propagated through three levels of beam-splitters.This probability distribution is different from the binomial distribution obtained in classical walks. We can see that even with a single photon, constructive and destructive interference still occurs. This interference is caused by the photon taking multiple paths to arrive at a specific detector.

\subsection{Twofold probability}

\begin{figure}[htbp]
  \centering
    \subfigure[] {\includegraphics[height=5.25cm]{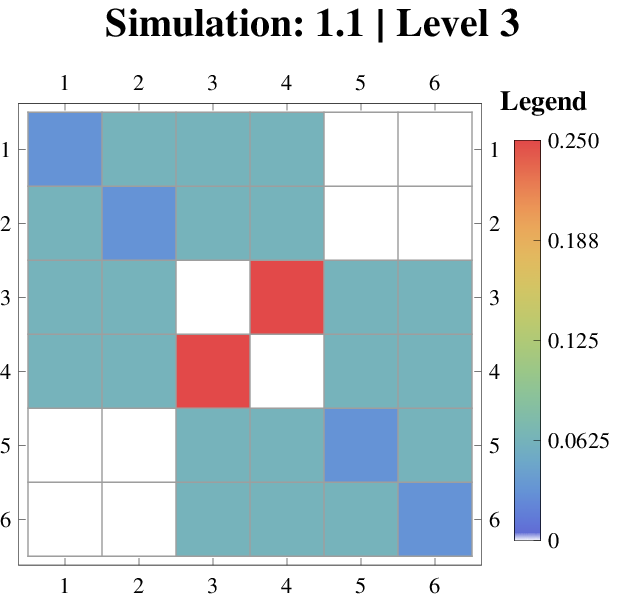}
    \label{fig:2foldplot}}
    \subfigure[] {\includegraphics[height=5.25cm]{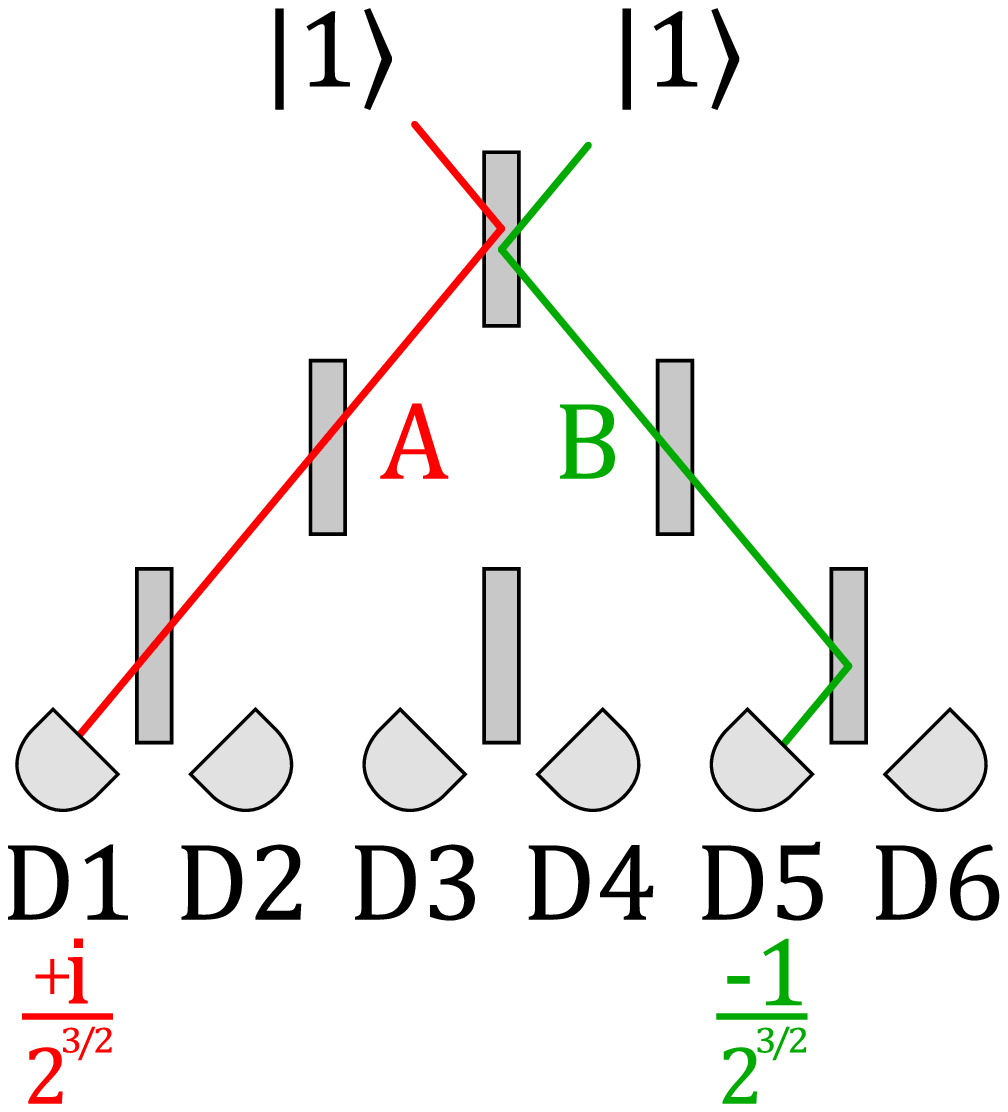} 
    \label{fig:2foldfeynmanL}}
    \subfigure[] {\includegraphics[height=5.25cm]{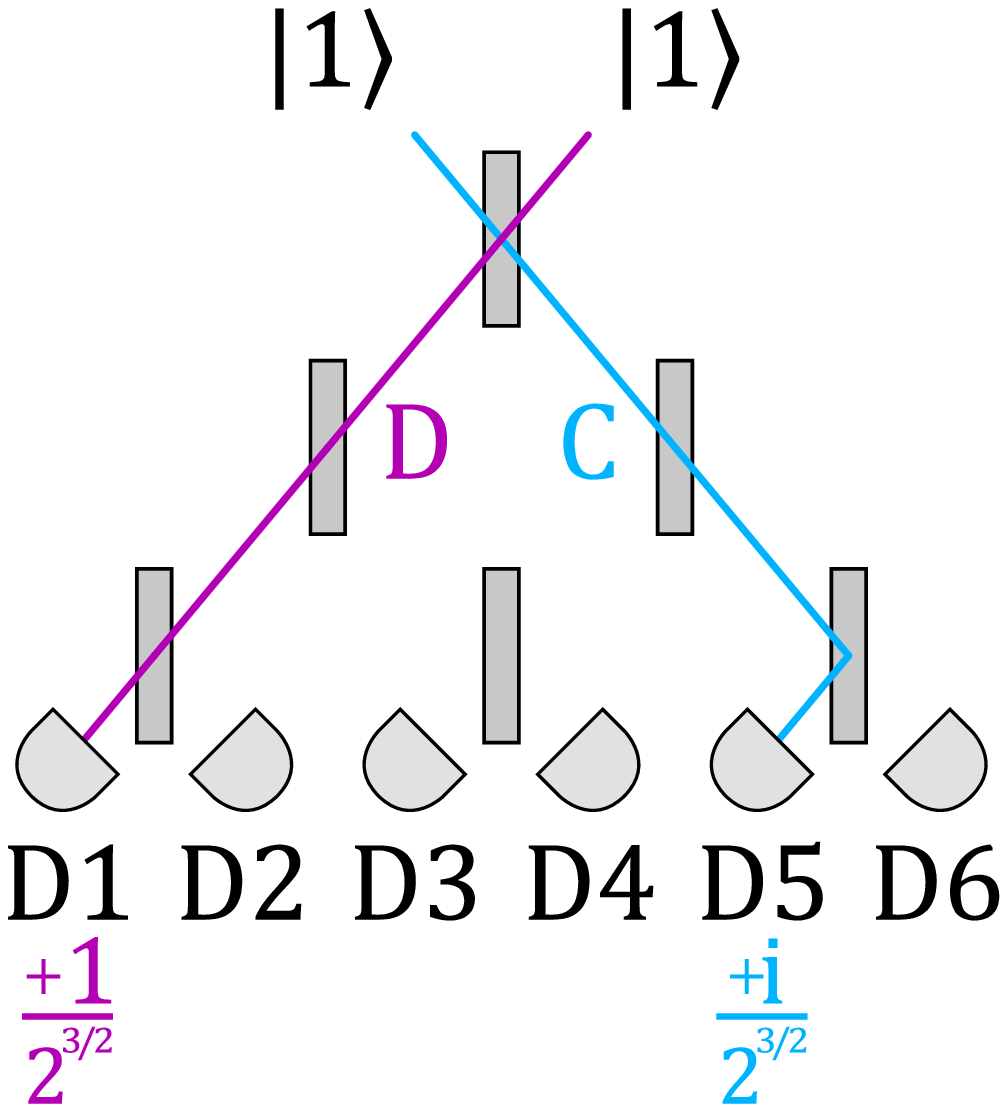}
    \label{fig:2foldfeynmanR}}
\caption{(Color online) (a) Plot showing the twofold probabilities using a $\ket{1,1}$ input state propagated through three levels. (b) Feynman diagram showing one set of unique paths that reach detectors D1 and D5. (c) Feynman diagram showing the only other unique set of paths that the photons can take to reach detectors D1 and D5.}
\end{figure}

Quantum random walks with multiple walkers require higher order correlation functions to reveal additional interference effects that take place. For a two walker random walk, $\ket{2,0}$ input state, along with its required normalization, produces results similar to $\ket{1,0}$ . Figure \ref{fig:2foldplot} shows a plot of our twofold probabilities for a $\ket{1,1}$ input state after three levels of propagation. The horizontal and vertical axes of the plot correspond to the two detectors $m$ and $n$. The symmetry can be seen down the diagonal of the plot.

\subsection{Threefold probability}
\begin{figure}[htbp]
  \centering
    \begin{minipage}{2in}
      \subfigure[] {\includegraphics[width=2in]{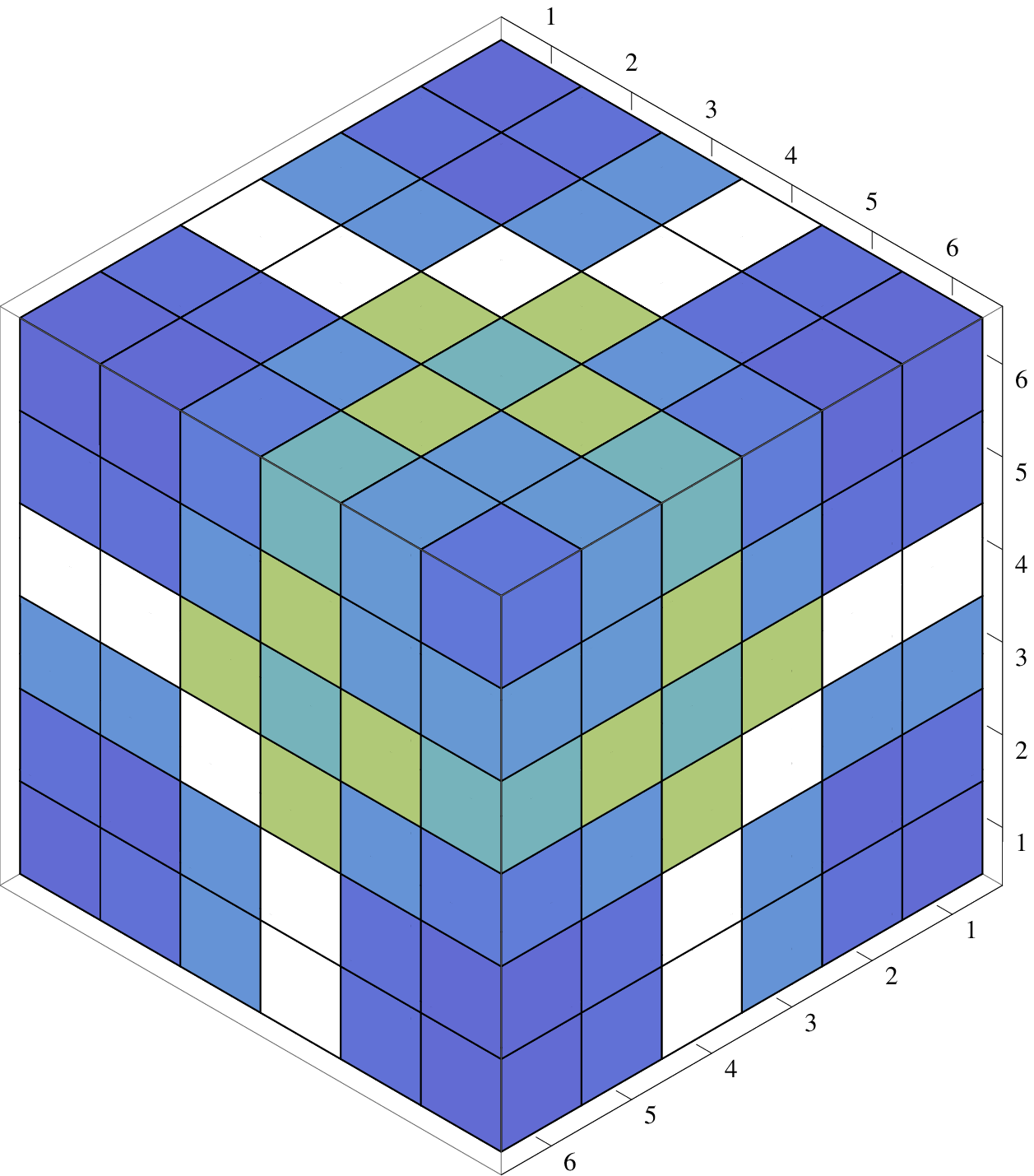}
      \label{fig:3fold-1}}\\
      \subfigure[] {\includegraphics[width=2in]{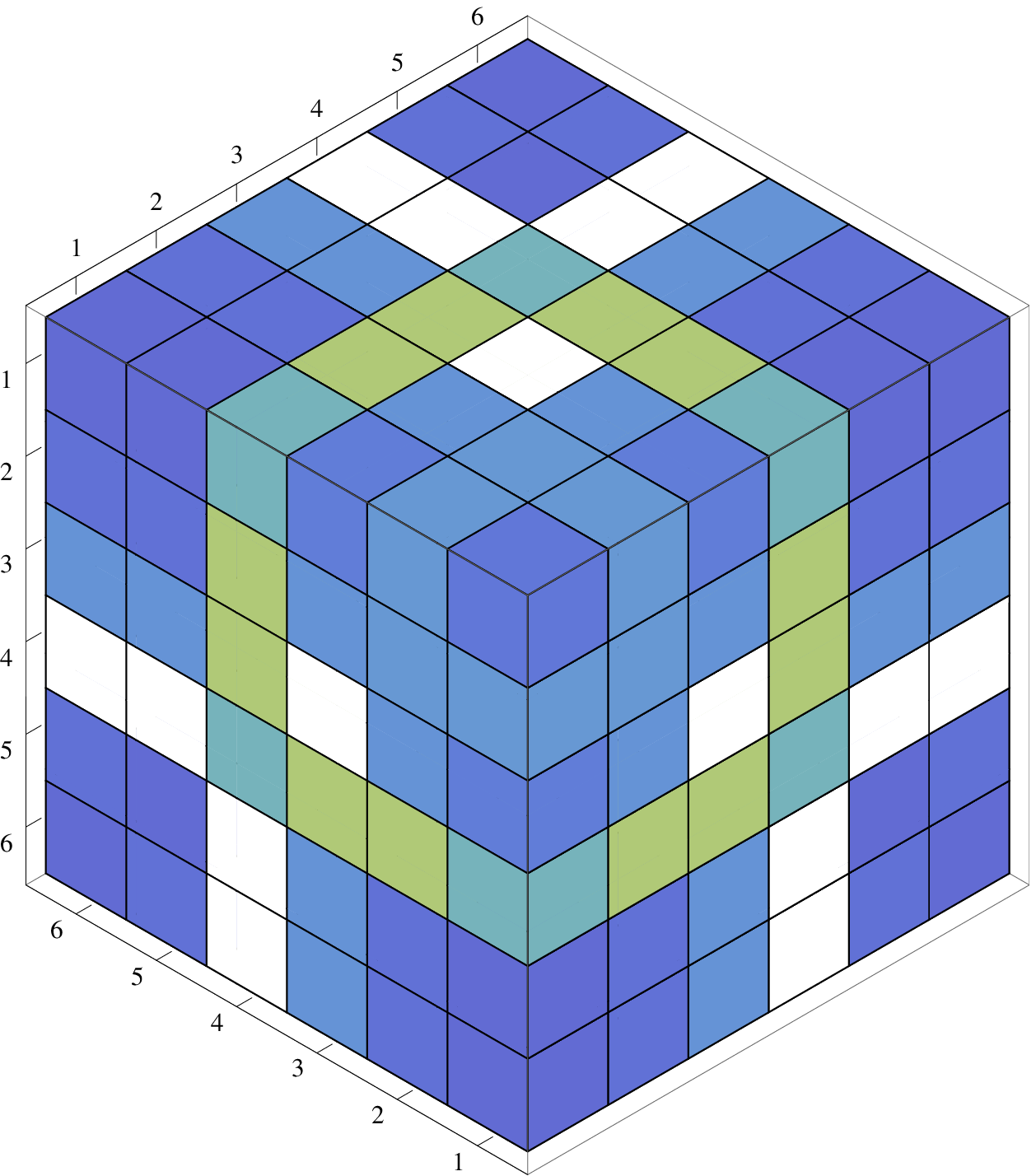}
      \label{fig:3fold-2}}
    \end{minipage}
    \hspace{.25cm}
    \begin{minipage}{2in}
      \subfigure[] {\includegraphics[width=2in]{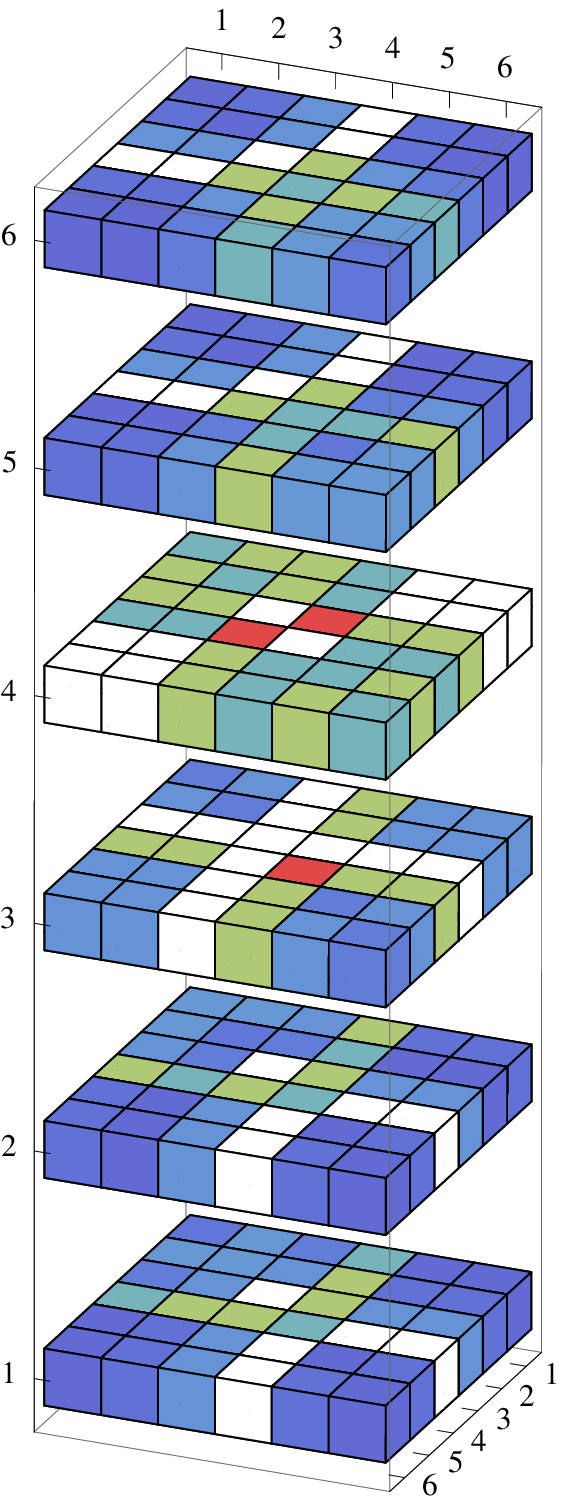}
      \label{fig:3fold-3}}
    \end{minipage}

\caption{(Color online) Plot of threefold probabilities using input state of $\ket{2,1}$ propagated through 3 levels of beam-splitters. Each row, column, and pillar represents an individual detector while the color of each cube represents the probability. Blue represents minimum probability, while red represents maximum, and white is exactly zero. (a, b) Plots of the threefold probabilities using input state of $\ket{2,1}$ propagated through 3 levels of beam-splitters detection scheme looking into the space diagonals. These views display the symmetry along the diagonals.  (c) ``Exploded" view of the same parameters. This view allows you to see inside the cube.}
\label{fig:3foldPlots}
\end{figure}

Increasing the number of walkers leads to higher order probability functions being deployed to reveal higher order interference effects that are a result of higher number of paths traveled by multiple walkers instead of individual paths traveled by a single walker. If we consider the case of three walkers, to visualize the results of the threefold probabilities on a specific input state, we must find some way of displaying probabilities at each of the three detectors independently. We overcome with this challenge by essentially stacking plots similar to our twofold probability plots in Figure \ref{fig:2foldplot}, each one keeping the third detector constant, and leaving space between each one in order to see into it. We call this the ``exploded Rubik's cube'' plot. Each of the three spatial dimensions are used to represent the three detectors $m$, $n$, and $l$, and the color of each square represents the threefold probability at those detectors. An example of this type of plot is shown in Figure \ref{fig:3fold-3}. Figure \ref{fig:3fold-1} and Figure \ref{fig:3fold-2} show the symmetry along the space diagonal that is present in all threefold probability plots. Using these plots we can also easily see interference effects that carry over from the onefold and twofold probabilities. For example, if for a specific input state and propagation level, the onefold probability at detector $a$ is zero, then for the same input state, the threefold probabilities at detectors $m$, $n$, and $l$ will equal zero for all $m, n, \textrm{or } l = a$. You can easily see this effect in the threefold probability plots, as entire planes will be zero. Similarly, if the twofold probability for a combination of detectors is, then an entire row and column in the threefold probabilities will be zero for that same combination of detectors.

\section{Discussion}
\subsection{Feynman Diagrams}
Coincidence detections with conventional photodiodes requires statistical accumulation of all possible outcomes in order to faithfully replace frequency of outcomes with probabilities. In particular, detection at $l^{th}$ level has $2l$ outputs. Therefore for the $N$-walker case there are $ (2l)^N$ outcomes that carry information about the quantum random walk. Each particular outcome provides only limited information about the quantum walk. In order to get an intuitive understanding of what is happening to the photons in quantum random walks, we use Feynman diagrams. We found that it is important to grasp how the phase shifts are being picked up and how they create constructive and destructive interference, visually. In the following sections, we will show specifically how the diagrams work, using examples of input states with a given number of photons.

\subsection{Onefold probability}
The complete destructive interference at detector D3 shown in Figure \ref{fig:1foldplot} distinguishes it from the classical random walk and can be explained using a Feynman diagram. Figure \ref{fig:1foldfeynman1} and Figure \ref{fig:1foldfeynman2} show the Feynman diagrams for the onefold probability of detector D3 firing with the $\ket{1,0}$ input state after three steps. These diagrams show all possible paths for the single photon to reach detector D3. In this case there are only two possible paths, which means that no matter what is happening to the photon on all other paths, this particular outcome is unaffected. Thus we have post-selected to show only those Feynman diagrams which contain paths that end at the desired detector.

In order to calculate the destructive interference, we define $t$ and $r$ as the coefficients that a photon's probability amplitude accumulates when it is transmitted or reflected.
\begin{eqnarray}
t &= &1/\sqrt 2,\qquad \textrm{(Photon is transmitted)} \nonumber \\
r &= & i /\sqrt 2, \qquad \textrm{(Photon is reflected)} 
\label{eq:trdef}
\end{eqnarray}
The definitions of $t$ and $r$ remain consistent with the beam-splitter transformation equation \ref{eq:creationtransformations}. It is important to note that these definitions for $t$ and $r$ do not include the required normalization factors for cases when the number of photons is greater than one.

When the photon follows path A in Figure \ref{fig:1foldfeynman1}, it is reflected off the first, second, and third beam-splitters. Therefore this path has a total probability amplitude of Path A = $r^3 = -\textrm i / 2^{3/2}$. When the photon follows path B in Figure \ref{fig:1foldfeynman2}, it is transmitted through the first beam-splitter, is reflected off the second, and is transmitted through the third beam-splitter. Therefore as the photon follows path B, it receives a total probability amplitude of Path B = $ t\,r\,t = \textrm i / 2^{3/2}$. In order to calculate the probability of photon detection at D3, the probability amplitudes are added together, Path A + Path B = $ -\rm i/ 2^{3/2} + \rm i/ 2^{3/2} = 0$. All the paths combine to zero, which confirms our result in Figure \ref{fig:1foldplot} at detector D3.

Constructive interference at D4 is revealed by looking at the Feynman diagrams in Figure \ref{fig:1foldfeynman3} and Figure \ref{fig:1foldfeynman4}, when we follow path A, we see that there are two reflections followed by a transmission. This gives us Path A =$ r^2\,t=-1/2^{3/2}$. Following path B, we see one transmission followed by two reflections. This gives us Path B =$ t\,r^2=-1/2^{3/2}$. In order to calculate the probability of photon detection at D4, we add these two probability amplitudes Path A + Path B =$ -1/\sqrt 2$. This is the probability amplitude of the photon being detected at detector D4. So then if we look at the probability $|-1/\sqrt 2|^2$ we see that the probability of the photon reaching detector D4 is $\frac{1}{2}$, as confirmed by our simulation in Figure \ref{fig:1foldplot}.

D3 and D4 present information about essentially the same path in the interferometer except for the final leg. However, from an experimental perspective, D3 is more informative since seeing a photon there tells us that something happened to the system. However, detecting or not detecting a photon at D4 does not immediately tell us that something has changed until one accumulates enough statistics.

\subsection{Twofold probability}

In order to obtain more information about the system, one needs to go to the higher order correlation function and multi walker cases. Let us consider destructive interference with two walkers. Using the $t$ and $r$ operators defined above, in equation \ref{eq:trdef} and referring to the Feynman diagrams in Figure \ref{fig:2foldfeynmanL} and Figure \ref{fig:2foldfeynmanR}, we calculate the twofold probability of detectors D1 and D5 firing simultaneously. Figure \ref{fig:2foldfeynmanL} shows one possibility of the two photons reaching detectors D1 and D5. These paths are described by

\begin{equation}
\textrm{Figure \ref{fig:2foldfeynmanL}} :
\left \{
\begin{array}{lr}
\textrm{Path A} = (r\,t^2)=i/2^{3/2} \\
\textrm{Path B} =(r\,t\,r)=-1/2^{3/2}
\end{array}  \right. .
\label{eq:2foldfeynmanL}
\end{equation}

Since these paths take place at the same time, we multiply these amplitudes together, $(i / 2^{3/2}) (-1 / 2^{3/2}) = -i/2^3$.
Figure \ref{fig:2foldfeynmanR} shows the only other paths the photons could take in order for detectors D1 and D5 to fire. These are described by

\begin{equation}
\textrm{Figure \ref{fig:2foldfeynmanR}} :
\left \{
\begin{array}{lr}
\textrm{Path C} =(t^2\,r)=i/2^{3/2} \\
\textrm{Path D} =(t^3)=1/2^{3/2}
\end{array}  \right. .
\label{eq:2foldfeynmanR}
\end{equation}

 Similarly, these two paths cannot take place independently of each other, so we multiply them, $(i / 2^{3/2}) (1 / 2^{3/2}) = i /2^3$. Adding two possible scenarios is necessary in order to calculate the probability of a joint event,  $(i / 2^{3/2}) ( -1 / 2^{3/2}) +(i / 2^{3/2}) ( 1 / 2^{3/2}) = -i /2^3 +i /2^3=0$. This complete destructive interference is shown in our plot of the twofold probability at detectors D1 and D5 in Figure \ref{fig:2foldplot}. One cannot predict this destructive interference from the onefold probability. It truly is multiphoton interference!
\subsection{Threefold probability}

\begin{figure}[htbp]
 \centering
   \subfigure[] {\includegraphics[height=5.25cm]{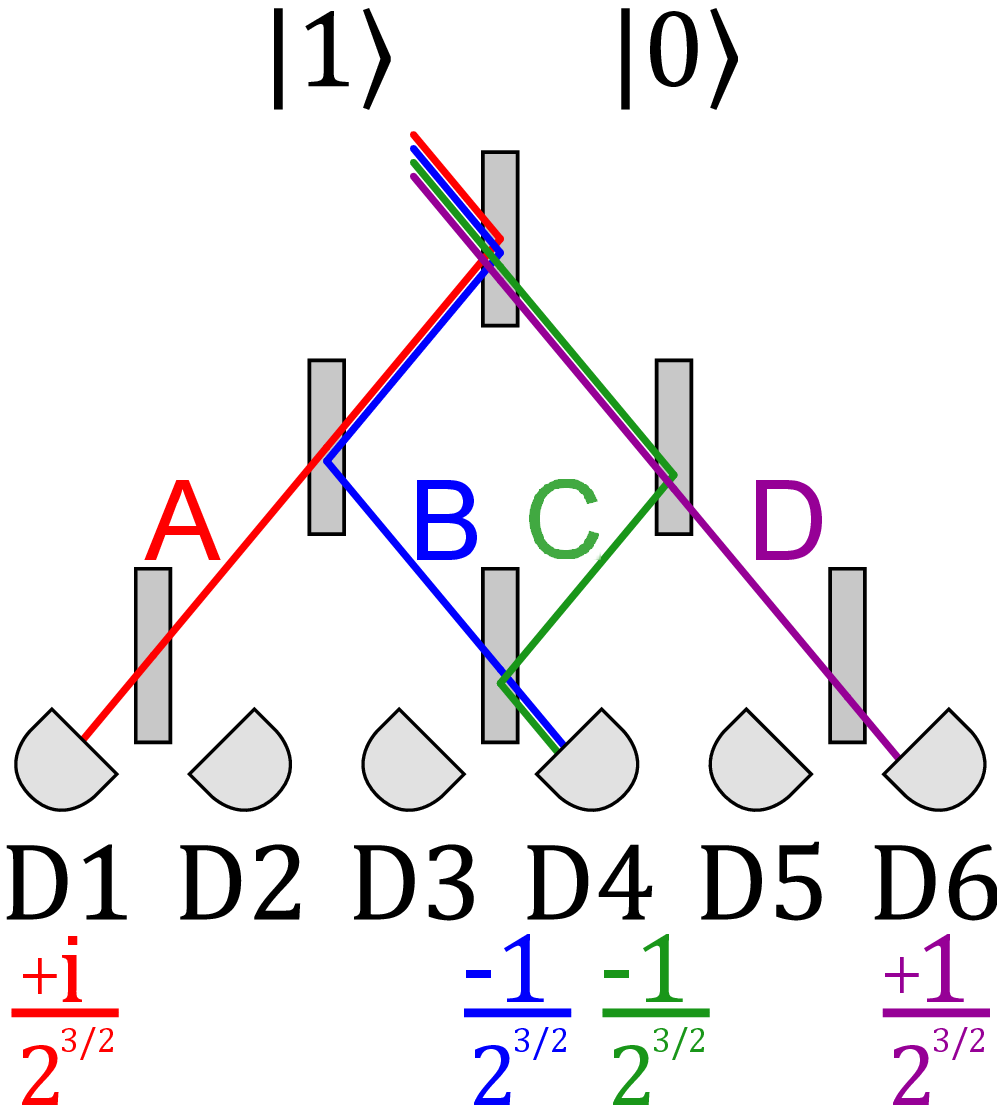}
    \label{fig:3foldfeynman1a}
    }
    \subfigure[] {\includegraphics[height=5.25cm]{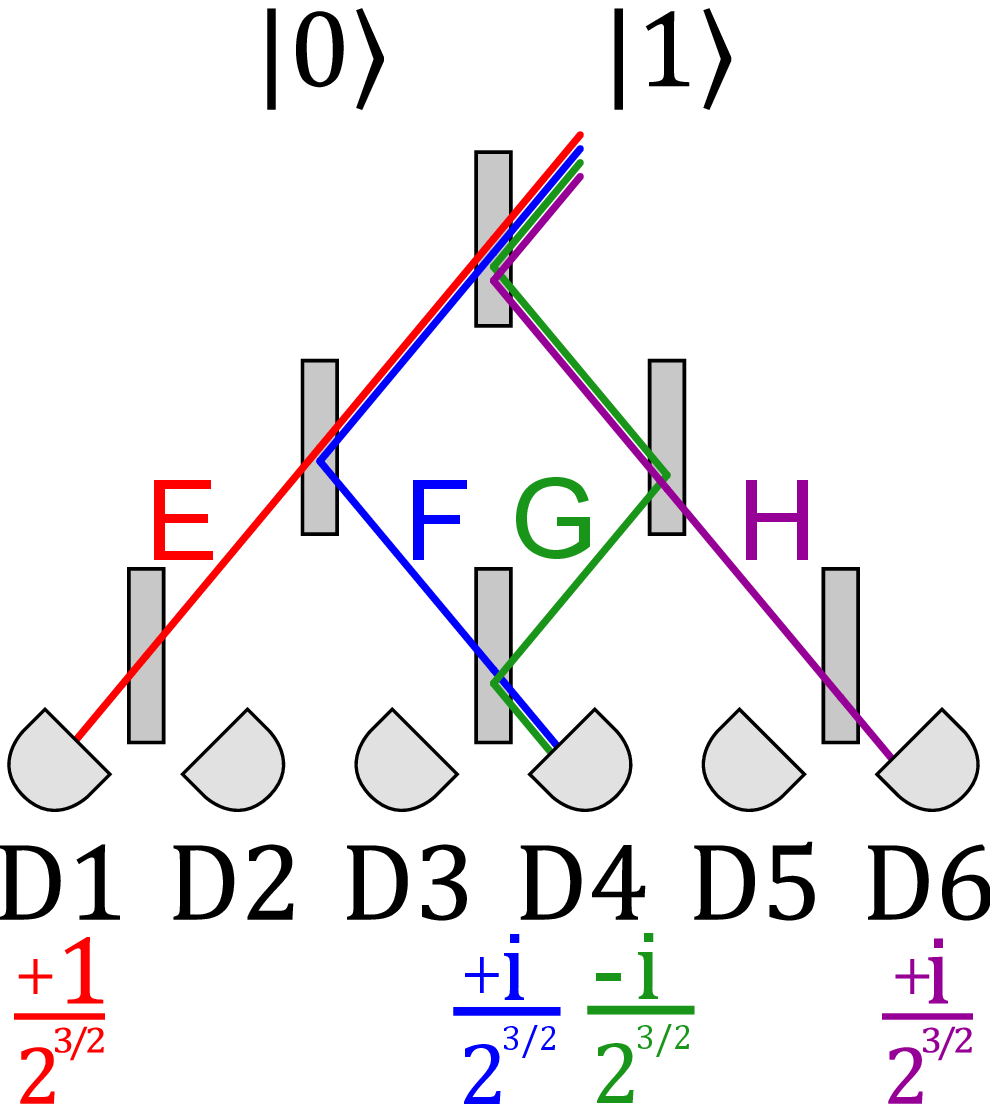}
     \label{fig:3foldfeynman1b}
     }
\centering
\caption{(Color online) Feynman diagrams explaining the threefold probability at detectors D1, D4, and D6 using the $\ket{1,0}$ and $\ket{0,1}$ as input states along with the discussion in section \ref{sec:model} to represent a $\ket{2,1}$ input state. These diagrams along with the discussion in section \ref{sec:model} show complete destructive interference with D1, D4, and D6 in coincidence. Note that each input state has two paths to arrive at detector D4 and thus has two amplitudes at D4.}
\label{fig:3foldfeynman}
\end{figure}

This detection scheme is more complicated than the previous detection schemes and thus the Feynman diagrams are also more complex. These diagrams become more complex because we have at least three photons in this setup but only two possible input ports. Because of this complication we use a different description of these diagrams by using the simplification discussed in Section \ref{sec:model} and thus can represent the input state of $\ket{2,1}$ in terms of the simpler input of $\ket{1,0}$ and $\ket{0,1}$. We call this method using ``simplified Feynman's''. We can thus analyze the simplified Feynman diagrams in Figure \ref{fig:3foldfeynman} just as we have done before.
 \begin{equation}
\textrm{Figure \ref{fig:3foldfeynman1a}} :
\left \{
\begin{array}{lr}
\textrm{Path A} = (r\,t^2)=i/2^{3/2} \\
\textrm{Path B} =(r^2\,t)=-1/2^{3/2} \\
\textrm{Path C} =(t\,r^2)=-1/2^{3/2} \\
\textrm{Path D} =(t^3)=1/2^{3/2} 
\end{array}  \right. .
\label{eq:3foldfeynman1eq1}
\end{equation}

Along with a similar analysis for Figure\ref{fig:3foldfeynman1b} and noting that the amplitude at detector D4 is simply the sum of the two amplitudes from both simultaneous paths. We now use this analysis  and realize that we need only use the appropriate powers ($N=2,M=1$) in order to get the desired analysis of the $\ket{2,1}$ input state. Using the amplitudes from Figure \ref{fig:3foldfeynman} along with the simplification discussed in Section \ref{sec:model} we see
\begin{equation}
 \ket{2,1}_{1,4,6}= \frac{1}{\sqrt{2!\, 1!}}\left( \left(\frac{i}{2^{3/2}}\right)_1 + \left(\frac{-2}{2^{3/2}}\right)_4+ \left(\frac{1}{2^{3/2}}\right)_6\right)^2 \left( \left(\frac{1}{2^{3/2}}\right)_1 + \left(0\right)_4 + \left(\frac{i}{2^{3/2}}\right)_6\right) 
\label{eq:threefoldsimple}
\end{equation}
where we have used subscripts to denote the amplitudes at each detector. Expanding this expression and post-selecting only the terms that coincide with D1, D4, and D6 coincidence we get
$$ \ket{2,1}_{1,4,6}=\left( \left(\frac{1}{2^3} \right)_{1,4,6} + \left(0\right)_{1,4,6} - \left(\frac{1}{2^3} \right)_{1,4,6} \right) =0 $$
So we expect to see complete destructive interference for the threefold probability involving detectors D1, D4, and D6. This is shown in Figure \ref{fig:3fold-3} as block $(1,4,6)=(4,1,6)=(6,1,4)=(1,6,4)=(4,6,1)=(6,4,1)$. Using this method of simplified Feynman's allows us to describe complicated input states without an exponential increase in the number of required diagrams but we also lose some of the intuitive nature of the diagrams as now we are using diagrams of a different input state instead of the one we are investigating.

\subsection{Rules for construction of Feynman diagrams}
Feynman diagrams allow us to describe specific scenarios within our QRW in terms of intuitive quantum interference. It is advantageous then to have a set of rules dictating how to generally construct these diagrams, based on the choice of, input state, detector combination, and level of propagation. We also discuss some of the implications of using Feynman diagrams.

Feynman diagrams grow exponentially more complicated with the level of propagation and number of photons, just as calculating these scenarios by hand or on computer would be, which might be expected as it is known that quantum random walks, at least in the case of using qubits and quantum coin flips, provide a framework for universal quantum computation. It is unknown if linear optical interferometers with arbitrary Fock state inputs also provides a path to universal quantum computing, but Aaronson and Arkhipov have recently provided evidence that such a scheme as ours is not efficiently simulatable on a classical computer and in fact can be used to solve certain classes of computationally hard problems such as computing the permanent of a large matrix \cite{aaronson}. This permanent finding algorithm has recently been demonstrated with a linear optical circuit in the group of White in collaboration with Aaronson and Ralph and others \cite{broome}. These works are suggestive that, in fact, linear optics plus 
large Fock state inputs is perhaps a road to universal quantum computation, in contradiction to a recent claim by Rohde \cite{rohde2}, but that remains an open question.

It can easily be shown that the number of paths for a given input state $\ket{N,M}$ and level $L$ is given by $2^{L(N+M)}$. This shows why we choose to use simplified Feynman diagrams in the case of $N+M>2$. For example, choosing $N = M = 9$ and $L = 16$ we have   $2^{288}=5 \times 10^{86}$ total possible paths, which is about four orders of magnitude larger then the number of atoms in the observable universe. So in the end, attempts to write down the exponentially large number of paths will be futile, but for low photon numbers and levels it is nevertheless a useful tool. This also prevents us from being able to write down a simple closed form for the output of an arbitrary input state and arbitrary level of propagation, or for that matter calculating it numerically, in agreement with the assertion of Aaronson and Arkhipov.

The main issue with using diagrams to describe our system is in complicated cases that have many interference terms. This is because when using diagrams we are essentially propagating interference effects through the entire system and combining their interference at the output. This process can sometimes lead to unnecessary paths or diagrams since at each path overlap, interference effects can occur that can lead to full cancellation of a path. As an example, if we look at Figure \ref{fig:2foldfeynmanL} and \ref{fig:2foldfeynmanR} which are the two ways detectors D1 and D5 can both receive a photon, we can see that these paths actually never occur due to the Hong-Ou-Mandel \cite{hom} effect that occurs at the first beam splitter. This HOM effect forces the output after the first beam splitter to have both photons in either the left or right output of the first beam splitter, but does not allow the photons to take separate paths. Due to this effect, it is impossible that the paths in these diagrams actually occur, but we do see that we correctly get full destructive interference at these pair of detectors. From this example we see that with our Feynman diagrams we gather all the interference effects at the end of our system, but these interference effects may have occurred earlier in the system and thus we may be considering scenarios that actually never occur due to early interference effects. So while some cases are more complicated than they need to be, we gain the advantage of a intuitive diagram to help show the path counting in our system. We also see that in the complicated case of the $\ket{2,1}$ input state we chose to use diagrams of a simpler case with the aid of simple expansion and post-selection. Without this simplification we would require twelve Feynman diagrams to fully describe this state with the previous methods used in the single and two photon coincidence cases.

The number of paths in each standard Feynman diagram is equal to the total number of photons in the system. The use of multiple diagrams for a single choice of input is then due to the fact that each photon is indistinguishable from another; therefore we must account for every photon reaching each of the desired detectors as separate cases. Using Figure \ref{fig:2foldfeynmanL} and Figure \ref{fig:2foldfeynmanR} as examples, there are two paths per diagram and two diagrams because there are two photons in this scheme but either photon can impact either of the detectors. Note that there are many other Feynman diagrams for the input of $\ket{1}$ in the left input and $\ket{1}$ in the right input, but we have post selected only the Feynman diagrams that have paths that end at detectors D1,D5.

For our simplified Feynman diagrams, we always use the states $\ket{1,0}$ and $\ket{0,1}$ along with the appropriate powers for arbitrary $\ket{N,M}$. The number of paths in each of these diagrams is then the number of paths that a single photon can take through the system to the desired detectors. It is clear that, in practice, this has its limitations as well as the number of paths for a single photon still increase exponentially with the level of propagation.

We define a pair of detectors as any two detectors that share the output of a common beam-splitter. The number of paths that lead to any pair of detectors are given by the binomial expansion,
\begin{eqnarray}
(x+y)^{(l-1)}=\sum\limits_{k=0}^{l-1} \left( 
\begin{array}{c}
l-1\\
k
\end{array} \right)
x^{(l-1-k)}y^k \nonumber
\end{eqnarray}
 where $l$ is the level of propagation and the coefficient of the $i^{th}$ term gives the number of paths for the $i^{th}$ pair of detectors. This shows that pairs of detectors that reside on the ``edge'' of the system only ever have one path that lead to them, regardless of level. It also shows that any ``non-edge'' detectors, which we call ``central-pair'' detectors, have multiple paths that lead to them. Using Figure \ref{fig:3foldfeynman1a} as an example, detector D4 is part of a central pair and thus has two paths that lead to it.

Another case that leads to multiple diagrams is when multiple photons are introduced from the same input port. This is always the case in the threefold detection scheme, as there are always three total photons but only two input ports. It is important to note that when using Feynman diagrams with multiple photons and a high order detection scheme, keeping track of all possible paths becomes much more cumbersome. Because of this reason, we turn to the use of the simplified Feynman diagrams for cases of threefold coincidence detection. This greatly reduces the required number of diagrams and also ensures proper normalization.

With these rules, in principle, one could use Feynman diagrams to describe any scenario within our system, but realistically, we limit the use of Feynman diagrams to describe low-level, few photon scenarios in order to reduce the number of required diagrams. The number of Feynman diagrams grows exponentially with increasing photon number and level as the associated Hilbert space needed to describe the system also grows exponentially with these parameters.

\section{Future research}

\subsection {Twofold number resolving detection}
With the recent development and use of number-resolving photon detectors, we propose a new detection scheme that can be utilized \cite{eisaman}. Our twofold-number resolving detection scheme is an extension of the twofold-detection scheme but with the added benefit of number resolving detectors. In this detection scheme we can investigate the probability of detector $m$ receiving exactly $i$ photons given detector $n$ receives exactly $j$ photons simultaneously. The result of this detection scheme would be four dimensional and would be much more difficult to visualize. We are currently analyzing the results of this detection scheme. The benefit of this scheme over the standard twofold scheme are the cases where there are more than two photons present in the system.

We are interested in investigating further modifications to the cascading beam-splitter structure. In our current simulation, we only introduce vacuum into the edge ports at each level. Introducing photons into the side ports in a systematic way may result in more interesting and useful output states. Another avenue could be the use of phase shifters in between specific beam-splitters. We could change the amount of relative  phase shift as well as the placement of these phase shifters. If placed properly, these phase shifters may effectively steer photons to a desired port with constructive interference or create a dark port by canceling out all phases at that detector. The possible combinations of placement and phase shifts would surely be enormous so we have considered implementing a genetic algorithm or similar technique to fine tune the system in order to create a desired output. We intend to look at beam splitters with other than 50-50 reflectivity-transmission where we will use this tuneablilty of the 
reflectivity as a means for programming the interferometer to simulate certain problems in quantum mechanics where an exponentially large state space is required.

\section{Conclusion}

We have shown that quantum random walks exhibit properties that are not present in their classical counterpart. Furthermore, multiphoton interference gives rise to new phenomena not found in standard quantum random walks.These properties are due to the quantum nature of the input photons and gives rise to unique interference effects. We simulate this quantum random walk with an array of 50/50 beam-splitters and a bank of detectors at any desired level of the system, only limited by the computational resources required to calculate these schemes. It has been suggested that these random walks have applications in quantum computing and solving the graph isomorphism problem  \cite{shenvi,lovett,gamble}.

We expanded upon the onefold probability detection scheme with our two and three fold coincidence detection schemes. Two fold detection has been studied previously but our three fold detection scheme, while harder to visualize, is the next logical step in simulating a quantum random walk and provides more information than a twofold probability scheme\cite{peruzzo}.

We use Feynman diagrams to give an intuitive explanation of how the interference effects propagate through our system and result in the final output distribution. These diagrams are functionally limited to low levels of propagation but still act as a valuable tool in order to explain how our simulation propagates the input photons.

\section*{Acknowledgments}
We would like to acknowledge Kaushik Seshadreesan, Barnabas Kim, Matthew Buras and Viv Kendon for encouragement and interesting discussions with this topic. This research was made possible through funding provided by AFOSR, FQXI, IARPA, NIST, NSF, and support by the LA-STEM program.

\end{document}